\documentstyle[prl,aps,twocolumn,epsf]{revtex}
\draft
\begin{document}
\twocolumn[\hsize\textwidth\columnwidth\hsize\csname @twocolumnfalse\endcsname

\title{Effect of Grain Geometry on Angle of Repose and Dynamics}
\author{C. J. Olson, C. Reichhardt, M. McCloskey, and R. J. Zieve}
\address{Department of Physics, University of California, Davis, California
95616.}

\date{\today}
\maketitle
\begin{abstract}
With simulation and experiment we examine the effect of grain anisotropy on
the angle of repose in a two-dimensional geometry.  We find that
the structure of the granular packing can be directly controlled by changing
the aspect ratio of the grains, and that the specific type of order   
strongly affects the angle of repose and the resulting dynamics.
The angle of repose is largest for dimer grains,
and is smaller for both monomer and trimer grains.
The motion of a collapsing grain heap consists of large scale collective
shear when monomer grains are considered, but occurs via tumbling of
grains on the surface for dimers, and via
disorderly bulk tumbling for trimer grains.
\end{abstract}
\vspace{-0.1in}
\vspace{-0.3in}
\vskip2pc]
\narrowtext

Granular assemblies have attracted much recent attention
since they are a model system in which
collective nonequilibrium dynamics, compaction, pattern
formation, frustration, nonlinear flows, and jamming 
can be readily examined \cite{JaegerReview1}. 
The behavior they display is relevant to a diverse array of
physical systems, including powders, colloids, liquids, and vortex matter 
in superconductors.  In addition, an understanding of granular
mechanics will be valuable for increasing the efficiency of a
wide range of industrial processes and applications.
A fundamental question that arises is the relation between
the mechanical response of the
granular media and the ordering of the individual grains.
The nature of the close-packed ordering in 3D of different types of
granular media is itself an open issue. 
In this work we study a simple system of granular assemblies confined in a
2D plane and show that specific types of ordering 
can be controllably produced by
tuning the anisotropy of the grains.  The resulting structure of the
granular pack has a profound effect on the angle of repose and the
granular dynamics.

Most previous work on the effects of grain geometry has focused on
angles of repose for grains that are roughly spherical.
The effect of dispersion in the sizes of individual spherical 
grains has been considered \cite{Cantelaube2}, 
as well as the effect of varying the roughness of
the grain surface \cite{Dury3}
or of having nonspherical grains 
\cite{Buchalter4,GallasI9}. 
Anisotropic or elongated grains have attracted interest previously, 
but it has generally been in the context of the 
flow of feed grains through a hopper \cite{hopper11}.
Only one systematic study has considered the effect of gradual elongation
of the grains on the angles of repose.  In Ref. 
\cite{GallasI9},
it was observed that as the grain is elongated continuously from a monomer
to a dimer, the angle of repose increases monotonically.  The grain
shape was continuously varied, so it would not be
possible to form an ordered lattice with the same lattice constant
using any two grain geometries considered.

In this work we consider with experiments and simulations a 
system composed of grains that
can always be packed into a triangular lattice of monomers,
but are individually composed of monomers, dimers, or trimers. 
Some previous simulation 
studies have focused on fully three-dimensional systems 
\cite{Baxter12,Zhou13}.
This introduces considerable complexity to the problem since there are
many more degrees of freedom, and 
it is difficult to 
determine experimentally the behavior of the grains below the surface of the
pile \cite{Drake13a}.  
For this reason, 
we consider a system 
constrained to move in two dimensions.  
Here the microscopic behavior
of individual grains can be directly observed experimentally and
compared with computer simulations.

With experiments and simulations we
find that the maximum angle of stability depends on the grain aspect ratio
in a non-monotonic way.  Both monomers, which are fully ordered,
and trimers, which are both positionally and bond orientationally
disordered, are less stable than dimers, which have positional order but
not bond-orientational order. 
We directly link these specific types of ordering to the 
grain dynamics, 
which 
range from orderly large scale shear flow for monomers,
to tumbling surface
flow for dimers, to disordered bulk flow for dimers. 
Our results in 2D represent a
first step towards the general understanding of how the 
microscopic structure affects the mechanical
response in 3D granular assemblies. 

\begin{figure}
\center{
\epsfxsize=3.5in
\epsfbox{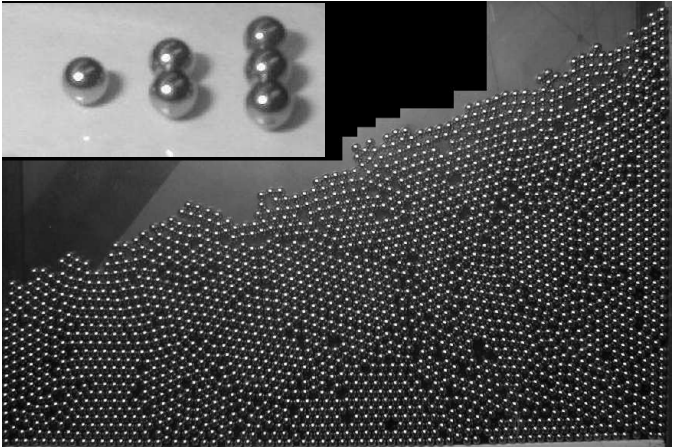}}
\caption{Digital camera image of experimental appearance of a system of dimers
after an avalanche has occurred.  Patches of order remain present
in the system.
Inset: Experimental monomers, dimers, and trimers.}
\label{fig:expt}
\end{figure}

{\it Experiment}---We consider a system of steel balls of radius
1/16'' constrained to move in 2D by
two Plexiglas sheets \cite{Rena14}.  
The steel balls are
either used individually, or are welded together to form dimers or
linear trimers (see inset to Fig.~\ref{fig:expt}).  All three shapes can
tile space in triangular arrays with the same lattice constant.
The grains are poured into a vertical enclosed area
9'' wide by 14'' high,  
and the container is briefly shaken vertically 
to prepare a more ordered initial starting condition.  Monomers and
dimers form nearly perfect triangular lattices under these conditions, 
but there is always a large amount of disorder present in the trimers  
\cite{Rena14}.
The container is slowly tilted in the plane of the Plexiglas sheets
to change the angle between the
surface of the granular packing and gravity \cite{Cantelaube2}.  
We record the angle at which the first grain motion occurs, giving us the 
maximum angle of stability $\Theta_s$ for the system, which is similar to the
angle of repose.  

In Fig.~\ref{fig:expt} we show the appearance of a system of dimers 
after 
an avalanche has occurred.
The grains remained motionless until 
the container had been
tipped by $60^{\circ}$, 

\hspace{-13pt}  
and then a large avalanche occurred.  A
significant amount
of order remains
in the system even after
the avalanche, 
as seen in Fig.~\ref{fig:expt}.
The maximum angles of stability $\Theta_s$ measured experimentally
in this way for grains of differing anisotropy
are listed in Table I.  We find the lowest angle $\Theta_s = 43^{\circ}$
for the disordered trimer system.  The well-ordered
monomers have a slightly higher value of $\Theta_s = 45^{\circ}$, which
is in reasonable agreement with previous experiments on 
spherical or nearly spherical grains \cite{Cantelaube2,Dury3}. 
In contrast dimers display a much larger angle of stability of 
$\Theta_s = 62^{\circ}$.
To better understand the reasons that
the dimer shape is so stable, we turn to a simulation of the system.

{\it Simulation}---We consider a system of $N_g=705$ grains constrained to
move in a 2D plane in a system of size $10 \times 18$ to
$60 \times 18$.  We use a granular dynamics method 
similar to the one employed in \cite{Gallas15,Herrmann16,Luding17} to integrate
the equations of motion for each grain, given by:
$m_i {\bf v}_i = {\bf f}_{\rm el}^{(i)} + {\bf f}_{\rm diss}^{(i)}
  + {\bf f}_{\rm shear}^{(i)} + {\bf f}_{g} + {\bf f}_{\rm fric}$.
Here, ${\bf f}_{g} = -0.0025$ is the force of gravity 
and ${\bf f}_{\rm fric} = 0.3$ is
the friction between the grain and the plane in which the grains move
(representing the Plexiglas plate).  
Two grains interact only when their relative distance is smaller than
the sum of their radii, $r_g = 0.4$.  
In this regime, three forces are active.
The elastic restoration force
between two grains is
${\bf f}_{\rm el}^{(i)} = \sum_{i\ne j} k_g m_{i}( |{\bf r}_{ij}| - 2r_g )
{\bf r}_{ij}/|{\bf r}_{ij}|$,
where $k_g=20$ is the strength of the restoring spring,
$2r_g$ is the grain diameter, $m_{i}=1$ is
the grain mass and ${\bf r}_{ij} = {\bf r}_{j} - {\bf r}_{i}$, the
distance between grains located at ${\bf r}_{i}$ and ${\bf r}_{j}$. 
The dissipation force due to the inelasticity of the collision is
${\bf f}_{\rm diss}^{(i)} = -\sum_{i \ne j}
\gamma m_{i}({\bf v}_{ij} \dot {\bf r}_{ij})
{\bf r}_{ij}/|{\bf r}_{ij}|^{2}$,
where $\gamma=2.4$ is a phenomenological dissipation coefficient and
${\bf v}_{ij} = {\bf v}_{i} - {\bf v}_{j}$ is the relative velocity.
The shear friction force mimicking solid friction is
${\bf f}_{\rm shear} = -\gamma_{s} m_{i}({\bf v}_{ij} \dot {\bf t}_{ij})
{\bf t}_{ij}/|{\bf r}_{ij}|^{2}$,
where $\gamma_s=1.2$ is the shear friction coefficient and
${\bf t}_{ij} = (-r_{ij}^{y},r_{ij}^{x})$ is the vector ${\bf r}_{ij}$ rotated
by $90^\circ$.  The behavior we observe is not sensitive to our choice
of parameters.

In this model, we neglect Coulomb friction and the rotation of individual 
particles.  In some cases, we constrain the grains
to form dimers or trimers by rigidly fixing the grains together.
These composite dimers and trimers are free to rotate.
The walls and floor are evenly lined with immobile grains of the same size as
the simulated monomers. 
Such a rough floor is necessary to produce a finite
angle of repose in the case of monomers; without it,
the monomers continue to slump until their height is reduced to
a single grain.  Dimers and trimers, however, 
produce a
finite angle of repose even on a smooth floor due to their tendency
to interlock. 

The system is prepared in one of two ways.
In studies of $\Theta_s$, the grains are prepared in
a box of size $10 \times 18$ such that the top surface of the grains
is roughly flat, 
either by dropping individual grains from above, or by placing them
into an ordered arrangement directly.
In studies of the dynamics of collapsing grains,
the grains are first introduced into a box
that is 1/6 as wide as the full simulation area of $60 \times 18$, 
either by dropping or by direct placement. 
The right wall is then instantaneously moved
out to the edge of the system, and the grains collapse to the right.
This method has been previously employed in Refs.~\cite{Elperin18,Lee19} to
study angles of repose.

\begin{figure}
\center{
\epsfxsize=3.5in
\epsfbox{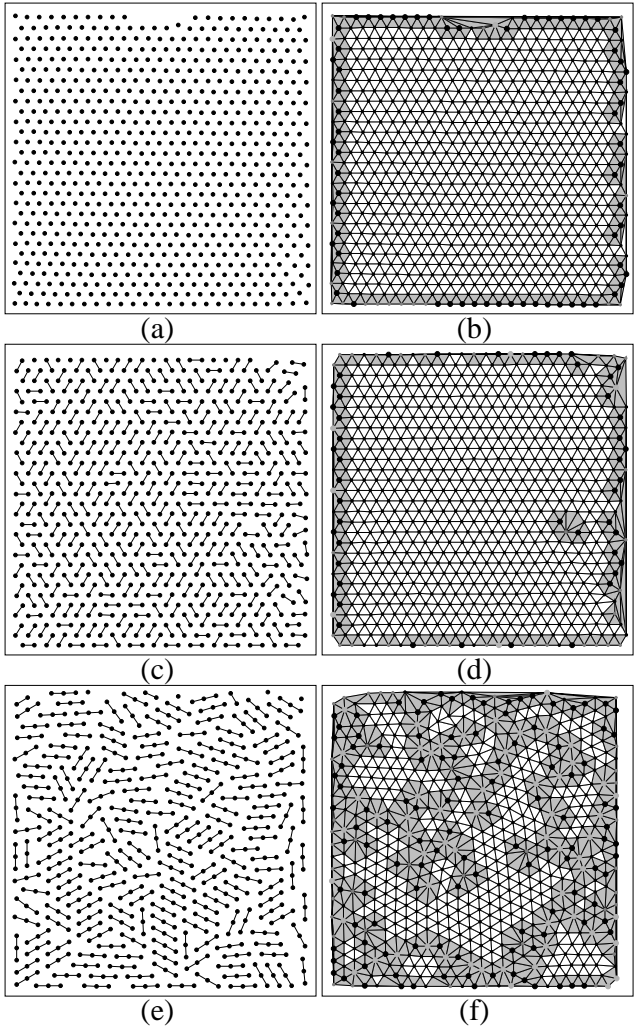}}
\caption{Static images from simulation after grains have been dropped
into a container.  Monomers: (a) positions, (b) Delaunay triangulation.
The system is well ordered.
Dimers: (c) positions, with lines connecting the dimer elements, (d)
Delaunay triangulation, with dislocation sites colored grey.  The positions
of the individual grains are ordered, but there is no order in the orientation
of the dimer bonds.
Trimers: (e) positions, with lines connecting the trimer elements, 
(f) Delaunay triangulation.  There is disorder in both position and bond
orientation.}
\label{fig:vorimage}
\end{figure}

When 
the grains are dropped slowly
from above, we find that monomers and
dimers fall into well-ordered arrangements but trimers are disordered.  
This is illustrated
in Fig.~\ref{fig:vorimage}, where the positions of the grains 
are shown in Fig.~\ref{fig:vorimage}(a,c,e), and the Delaunay 
triangulations of the lattice
are shown in Fig.~\ref{fig:vorimage}(b,d,f).  The monomers of 
Fig.~\ref{fig:vorimage}(a,b) are perfectly ordered, with no
dislocations, as indicated in Table I.  For dimers as in 
Fig.~\ref{fig:vorimage}(c,d) the positional order is nearly 
complete but there is a slight chance for a vacancy to occur. 
Note that although the individual elements composing
the dimers are ordered, the bond angles have no overall order.
For trimers, shown in Fig.~\ref{fig:vorimage}(e,f), there is some
local ordering of the bond angles, but no overall order, and there is a
large amount of positional disorder in the lattice.

There is a simple geometric reason
for the difference in packing of the three grain shapes.  
If a layer of previously dropped grains is present, and there is
a single monomer-sized vacancy in the layer, 
then an additional monomer 
or dimer dropped over the vacancy will fall 
into the vacancy and fill it.
The dimer is not long enough to span the 
vacancy and one side of it will
slip in.  In contrast, the trimer is long enough to span the vacancy and
block it from being filled.  Additionally, a trimer is long enough to
assume a continuum of angles if dropped near a second trimer lying flat,
depending on the exact relative spacing 
of the two trimers,
whereas a dimer will assume only the angles allowed by a triangular
lattice. 
Thus with trimers only very local ordering is
possible and long-range order cannot be achieved through random dropping.

We next consider $\Theta_s$,
measured with the same 

\hspace{-13pt}  
procedure followed in the experiment.  
We simulate tipping
the box slowly by
rotating the 
direction of gravity, and 
record the angle $\Theta_s$
at which the first 
avalanche occurs.
As listed in Table I, we find
the same trend as experiment: low values of $\Theta_s = 27^{\circ}$ 
for ordered monomers and $\Theta_s = 13^{\circ}$
for disordered
trimers, and a much larger angle of $\Theta_s = 40^{\circ}$ 
for the dimers.
It is clear that the different types of order in these three geometries
lead to the different angles of stability.  The disordered trimers are
very unstable to 
rearrangements since there are many dislocations in
the granular packing where motion can occur.  The ordered monomers are more
stable due to

\begin{figure}
\center{
\epsfxsize=3.5in
\epsfbox{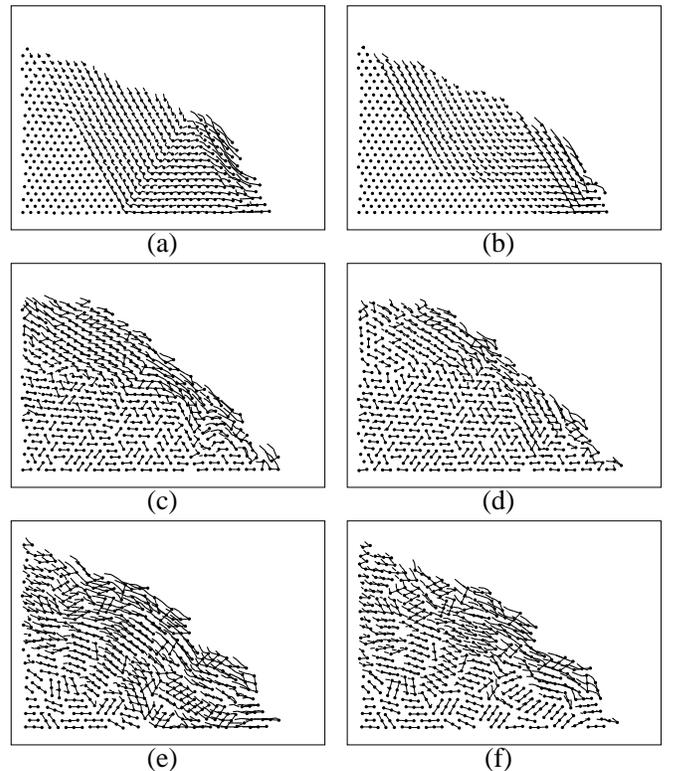}}
\caption{Dynamic images of granular collapse from simulation 
at pairs of consecutive time intervals after the
right wall of the container has been removed.  Circles indicate the 
location of individual grain elements, light lines indicate bonds for
dimers and trimers, and heavier lines indicate the trajectories over a
short time of the individual grain elements.  (a,b) Monomers, showing
orderly flow along lattice vector directions of large triangular wedges.
(c,d) Dimers, showing tumbling flow along the top surface of the pile
only.  (e,f) Trimers, showing disorderly flow throughout the bulk,
with tumbling motion at the top of the pile.}
\label{fig:image}
\end{figure}

\hspace{-13pt} 
the absence of dislocations, but are still unstable to shearing
motion along the lattice directions.  
In contrast, the dimers
are stabilized by the absence of dislocations like the monomers, but
shearing motion is suppressed due to the randomly arranged bonds
between the dimers, which lead to interlocking.  The larger $\Theta_s$
we observe for dimers compared to monomers agrees with the results of 
Ref.~\cite{GallasI9}.  Our study of the even more anisotropic 
trimers indicates, however, 
that
the increase in $\Theta_s$ with anisotropy found in Ref.~\cite{GallasI9}
does not continue indefinitely but instead decreases again as the
grains become longer than dimers.

To further explore how the structure of the granular packing affects its
stability, we directly examine 
the 
dynamics of collapsing grains when one wall is removed, and
compare the three geometries in Fig.~\ref{fig:image}.
We see a striking difference in the dynamics.  
Collapsing monomers move in large collective motions involving 
much
of the bead pack, which follow the lattice vectors.  The motion is very
orderly and occurs in the form of large pulses, during which triangular
wedges of grains displace along their 

\begin{figure}
\center{
\epsfxsize=3.5in
\epsfbox{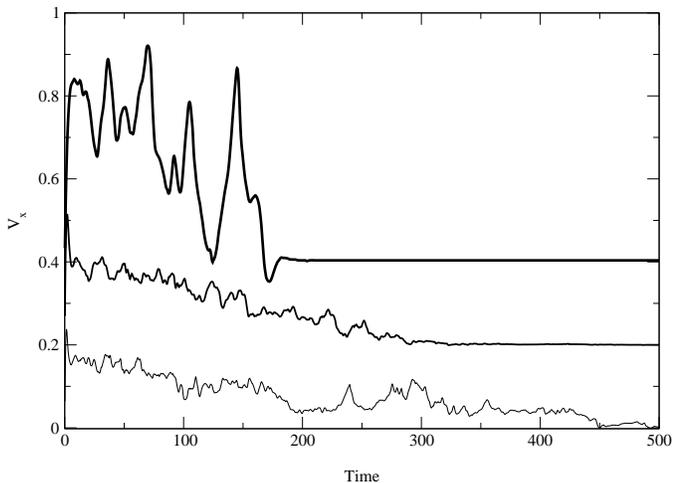}}
\caption{Net grain velocity in the $x$ direction, $V_x$, 
from simulation.  Top (heavy) line: Monomers, showing
large collective oscillations.
Middle line: Disordered dimers. Bottom (light) line: Trimers.  The curve
for the monomers has been vertically offset by 0.4 and the curve for dimers
has been offset by 0.2 for clarity.}
\label{fig:time}
\end{figure}

\hspace{-13pt} 
lattice vectors.
This wedge-like 
motion is similar
to 
motion 
seen in experiments with 
flowing steel
balls \cite{Berg20}.

In collapsing
dimers, the motion is limited to the surface of the pile due to the
interlocking of the dimers which {\it prevents} shearing motion along the
lattice vectors.  The motion that does occur is tumbling in nature and
very disorderly.  A relatively small portion of the bead pack is moving
at any given time.

In collapsing trimers, the large number of
dislocations present in the system serve as ``lubrication'' to the
motion, allowing easy disorderly sliding of a large portion of the
bead pack.  There is no significant ordering present on a scale of more
than a few grains so coherent motion is impossible.  Motion occurs
through a combination of gliding along the dislocated area, and tumbling
down the topmost portion of the pile.
Although the motion is widespread, it occurs more slowly than the
collective motion seen in the monomers.  This is due to the fact that 
a considerable amount of rearrangement within the bulk must occur as
the movement continues, unlike the case of monomers when no rearrangement
was necessary and the movement could follow the lattice vectors.

The different dynamics can also be observed by measuring the net
velocity in the x-direction, $V_x = \sum v_x$, of the grains over
time, as shown in Fig.~\ref{fig:time}.  Monomers produce extremely
large pulses as the collective motion occurs, whereas dimers and
trimers produce much smaller amplitude signals with their more
disperse responses.

{\it Summary}---
We have examined the ordering and dynamics of monomer, dimer and trimer
assemblies of grains confined in 2D and relate the specific types of ordering
to the dynamics of granular collapse.
Dimers, which have translational order but not bond-orientational order, 
show the largest angle of repose.  Monomers, which are completely 
ordered, show a smaller angle of repose that is still higher than
trimers which are completely disordered. 
The dynamics of the monomers 
during collapse occurs by large scale collective shear motions.  
The dimers, which do not allow for easy shear motion  
show a surface tumbling motion. Trimers also show a tumbling motion that
can occur in the bulk as well as on the surface.  

Acknowledgements: We thank H. Jaeger and N. Gr{\o}nbech-Jensen 
for helpful discussions.  This work
was supported in part by CLC and CULAR (LANL/UC) and by
NSF, DMR-9733898.

\begin{table}
\begin{tabular} {c|cccc}\hline
{Grain type} & {$\Theta_{s}$ (Expt.)} & {$\Theta_{s}$ (Sim.)} 
& {$P_{d}$ (Expt.)} & {$P_{d}$ (Sim.)} \\ \hline
{Monomer} & {$45^\circ$} & {$27^\circ$} & { 0\% } & { 0\% }\\ 
{Dimer} & {$62^\circ$} & {$40^\circ$} & { 3\% } & { 3\%} \\
{Trimer} & {$43^\circ$} & {$16^\circ$} & { 22\%} & { 30\%} \\ 
\end{tabular}
\caption{Table of results from experiment and simulation.  The angle
of stability $\Theta_{s}$ 
was obtained by tipping the container and identifying the
first motion of the flat granular surface.  $P_d$ 
is the percentage of individual grain elements that do not have
six neighbors (0 indicates perfect order).}
\end{table}

\end{document}